\documentclass[preprint,aps,prd]{revtex4}
\RequirePackage{xspace}

\usepackage{graphicx}
\def\jpsi{\ensuremath{{J\mskip -3mu/\mskip -2mu\psi\mskip 2mu}}}
\def\Kstarsz{\ensuremath{K^*_0}}
\def\Kstart{\ensuremath{K^*_2}}
\def\KstarX{\ensuremath{K^*_X}}\def\Kstar{\ensuremath{{K^*}}}

\def\beq{\begin{equation}}        
\def\eeq{\end{equation}}
\def\bea{\begin{eqnarray}}
\def\eea{\end{eqnarray}}
\def\Im{{\rm Im}}
\def\Re{{\rm Re}}
\def\nn{\nonumber}
\def\sss{\scriptscriptstyle}

\def\bd{B_d^0}
\def\bdbar{{\overline{B_d^0}}}

\def\barp{{\raise.35ex\hbox
{${\sss (}$}}---{\raise.35ex\hbox{${\sss )}$}}}
\def\bdbarp{\hbox{$B_d$\kern-1.4em\raise1.4ex\hbox{\barp}}}
\def\bsbarp{\hbox{$B_s$\kern-1.4em\raise1.4ex\hbox{\barp}}}
\def\ks{K_{\sss S}}

\def\roughly#1{\mathrel{\raise.3ex\hbox
{$#1$\kern-.75em\lower1ex\hbox{$\sim$}}}}

\def\adir00{{a_{\sss dir}^{00}}}

\def\B00{B^{00}}
\def\Bp0{B^{+0}}

\def\dsp{\displaystyle}

\newcommand{\tev}{\ensuremath{\mathrm{Te\kern -0.1em V}}\xspace}
\newcommand{\gev}{\ensuremath{\mathrm{Ge\kern -0.1em V}}\xspace}
\newcommand{\mev}{\ensuremath{\mathrm{Me\kern -0.1em V}}\xspace}
\newcommand{\kev}{\ensuremath{\mathrm{ke\kern -0.1em V}}\xspace}
\newcommand{\ev}{\ensuremath{\mathrm{e\kern -0.1em V}}\xspace}
\newcommand{\gevc}{\ensuremath{{\mathrm{Ge\kern -0.1em V\!/}c}}\xspace}
\newcommand{\mevc}{\ensuremath{{\mathrm{Me\kern -0.1em V\!/}c}}\xspace}
\newcommand{\gevcc}{\ensuremath{{\mathrm{Ge\kern -0.1em V\!/}c^2}}\xspace}
\newcommand{\mevcc}{\ensuremath{{\mathrm{Me\kern -0.1em V\!/}c^2}}\xspace}

\begin{document}

\preprint{IMSc-2005/02/02}

\title{Angular analysis of B decaying into \jpsi Tensor, \jpsi Vector
  and \jpsi Scalar
  modes}\author{Chandradew Sharma} \email{sharma@imsc.res.in} 
\author{Rahul Sinha}\email{sinha@imsc.res.in} 
 \affiliation{The
  Institute of Mathematical Sciences, C.I.T. Campus, Taramani, Chennai
  600113, India.}  
\date{\today}
\begin{abstract}
  The analysis of $B \to \jpsi K^{\star}_{2}(1430)$ decay mode is
  complicated by the fact that close to the $J^{PC}=2^{++}$ meson
  $\Kstart(1430)$, there lie other $J^{PC}=1^{--}$ and $J^{PC}=0^{++}$
  resonances, $K^{\star}(1410)$ and $K^{\star}_{0}(1430)$
  respectively.  We show how an angular analysis can be used to
  isolate the contributions from the different resonances and partial
  waves contributing to the final state $B\to \jpsi K_X$, where $K_X$
  could be any of the resonance $K^{\star}_{2}(1430)$,
  $K^{\star}(1410)$ or $K^{\star}_{0}(1430)$. For this purpose we
  study the time integrated differential decay rate. We also construct
  a time dependent angular asymmetry that enables a clean measurement
  of the mixing phase $\beta$ in the mode $B\to \jpsi \Kstart(1430)$
  alone, without contributions from the decay modes
  $B\to\jpsi\Kstarsz(1430)$ or $B\to\jpsi\Kstar(1410)$.

\end{abstract}

\maketitle

\section{Introduction}
\label{sec:1}

The large numbers of B mesons produced at the B-factories have resulted
in an accurate measurement\cite{Abe:2004mz} of the $\bd-\bdbar$ mixing phase refereed to
as $\beta$ (also called $\phi_1$). This measurement is done primarily using the
golden mode $B\to \jpsi\ks$, but is now being done for a variety of
other channels. In principle, $\beta(\phi_1)$ can be measured just as
cleanly using any mode that involves the same quark level process
$b\to c\bar{c}s$. These include modes that involve excitations of the
$K$ meson such as $B\to \jpsi \Kstar(892)$, $B\to
\jpsi\Kstarsz(1430)$, $B\to\jpsi\Kstar(1410)$, 
$B\to \jpsi
\Kstart(1430)$, etc., or modes involving various excitations of the
$c\bar{c}$ mesons such as $B\to \jpsi ^\prime \ks$ etc., or even a
combination of any of these states.

The study presented in this paper is inspired by the recent
observation of the decays $B\to \jpsi \KstarX(1430)$, where $X=0,2$
\cite{Hagiwara:2002fs,Coan:1999kh,Aubert:2001pe}.  There exist two
mesons, a tensor meson $\Kstart(1430)$ and a scalar meson
$\Kstarsz(1430)$ at the same mass of $1430$ \mev.  One is not only
interested in obtaining the $\bd-\bdbar$ mixing phase precisely, but
also in measuring the branching ratios for the various decay modes.
The branching ratio for $B\to \jpsi\Kstar$ has been measured and it is
well known that $\beta(\phi_1)$ can also be measured using this decay
mode\cite{Review}; the only complication is the need for an angular
analysis to separate the contributions from the $CP$--even and
$CP$--odd partial waves. However, modes such as $B\to\jpsi
\Kstart(1430)$ require much more effort.  The additional complication
arises due to the presence of scalar and vector meson resonances
$\Kstarsz(1430)$ and $\Kstar(1410)$ that overlap with the tensor meson
$\Kstart(1430)$.  Since, the decay modes $B\to \jpsi\Kstarsz(1430)$,
$B\to\jpsi\Kstar(1410)$ and $B\to \jpsi \Kstart(1430)$ contribute to
the same final state $B\to K\pi\ell^+\ell^-$, contributions from the
various decay channels cannot be separated by cuts on the kinematics.
The purpose of the paper is to show how the contributions from various
resonances and partial waves can be isolated, not only to measure the
branching ratios but also to study the time dependent decay to the
mode $B\to \jpsi \Kstart(1430)$, leading to a measurement of
$\sin2\beta$. The study
performed here finds immediate application in the analysis of data
collected by BaBar and Belle collaborations at the B factories running
at SLAC (U.S.A.) and KEK (Japan) respectively \cite{Review}.
 
In section~\ref{sec2} we write out the most general effective matrix
element using Lorentz invariance and current conservation, for
the decay channels $B\to\jpsi S$, $B\to \jpsi V$ and $B\to \jpsi T$,
where $S$, $V$ and $T$ are scalar ($J^{P}=0^{+}$), vector
($J^{P}=1^{-}$) and tensor ($J^{P}=2^{+}$) mesons respectively.  Our
calculations are general enough in formalism to include any scalar,
vector or tensor meson, in addition to $\KstarX(1430)$ and
$K_1(1410)$, as long as they decay into modes identical to the
modes into which $\KstarX(1430)$ is reconstructed.  The $\KstarX(1430)$ and
$K_1(1410)$ are considered to decay to $K\pi$ and $\jpsi$ to
$\ell^+\ell^-$. In section~\ref{sec3} we primarily investigate the
decay spectrum for the final state $K\pi\ell^+\ell^-$ assuming that
the $K\pi$ arise from the decay of either $S$, $V$ or $T$.  We study
the angular distribution of the $K$ in the $K\pi$ center of mass
(c.m.) frame and $\ell^-$ in the $\ell^+\ell^-$ c.m. frame. We also
study the correlation between the $K\pi$ decay plane and the
$\ell^+\ell^-$ decay plane in the B rest frame. We explicitly
demonstrate how one can isolate the contributions to the different
final states as well to $CP$ even and $CP$ odd partial waves, thereby
allowing the measurement of $\beta(\phi_1)$. We conclude in
Sec.~\ref{sec4}.

\section{Matrix Element}
\label{sec2}

We consider exclusive two body decays of a B meson into states
involving the \jpsi or its excitations (all these states will be
generically referred to as \jpsi). In particular, we restrict
ourselves to the decays $B\to\jpsi S$, $B\to \jpsi V$ and $B\to \jpsi
T$, where $S$, $V$ and $T$ are scalar ($J^{P}=0^{+}$), vector
($J^{P}=1^{-}$) and tensor ($J^{P}=2^{+}$) mesons respectively. We
assume that each of the resonances $S$, $V$ and $T$ decay into $K\pi$,
and that \jpsi is reconstructed in the $\ell^+\ell^-$ decay mode.
Hence each of the three decay channels results in the same decay
process $B \rightarrow K\pi\ell^{+}\ell^{-}$, allowing for
interference between the three channels.  This decay involving a 4
body final state, can be described in terms of 5 variables $s_{\ell}$,
$s_{K}$, $\theta_{\ell}$, $\theta_{K}$ and $\phi$. The kinematical
variables $s_{l}$ and $s_{K}$ are the invariant mass squared of the
lepton pairs($\ell^+ \ell^-$) and the $K\pi$ pairs respectively (it is
assumed that the $\ell^{+} \ell^{-}$ momentum is along the $+z$ axis),
$\theta_{l}$ is the angle of $ \ell^{-}$ in the $\ell^{+}\ell^{-}$
c.m. system with the $z$-axis, $\theta_{K}$ is the angle of $K$ in the
$K\pi$ c.m. system with the $z$-axis, and $\phi$ is the angle between
the normals to the planes defined by momenta of $\ell^{+}\ell^{-}$ and
$K\pi$, in the $B$ rest frame\cite{Kruger:1999xa}. The 4-momenta of
$S$ (or $V$ or $T$) is assumed to be $k$ and that of the \jpsi to be
$q$.  The $K$, $\pi$, $\ell^-$ and $\ell^+$ are defined to have the
4-momenta $k_1$, $k_2$, $q_1$ and $q_2$ respectively. We further
define $K^\mu=k_1^\mu-k_2^\mu$ and $Q^\mu=q_1^\mu-q_2^\mu$ and note
that $s_l=q^2$ and $s_K=k^2$.

Let us first consider the decay $B\to \jpsi S$. The most general
amplitude for this decay mode may be written using Lorentz
invariance as:
\begin{equation}
  \label{eq:amp-S}
   {\cal A}(B\to \jpsi(q) S(k))=
a k_\mu \epsilon_\jpsi^{*\mu}~,
\end{equation}
where $\epsilon_\jpsi^*$ is the polarization 4-vector of the $\jpsi$, $q$ and $k$ are the 4-momentum of the $\jpsi$ and $S$ respectively, and
$a$ is Lorentz scalar.  
The subsequent decay of the scalar S into two pseudoscalars, i.e.
$S\to K \pi$, will only result in the multiplication of the above
amplitude by an arbitrary function of $k_1.k_2$.  Therefore, the
amplitude for the process $B\to \jpsi S\to \jpsi(K \pi)_S$ may be
written as:
\begin{eqnarray}
  {\cal A}_S & = &{\cal A} \bigg( B \to \jpsi (K \pi)_S \bigg) 
    \propto a k_\mu \frac{1}{(k^2-M^2_S+i\epsilon)}
   \epsilon_\jpsi^{*\mu}~,
\end{eqnarray}
where, the subscript $S$ in $(K \pi)_S$ indicates that the $K\pi$ state
results from the decay of $S$. The amplitude for$B\to \jpsi S$
consists of $P$ wave only ~.

The amplitude for $B\to \jpsi V$ may similarly be written as,
\begin{equation}
  \label{eq:amp-V}
{\cal A}(B\to \jpsi(q) V(k))\propto (b g_{\mu\nu} +\frac{c}{\sqrt{s_{\ell}s_K}}k_\mu q_\nu + i \frac{d}{\sqrt{s_{\ell}s_K}}\epsilon_{\mu\nu\alpha\beta} k^\alpha q^\beta)\epsilon^{*\mu}_\jpsi\epsilon_V^{*\nu}~,
\end{equation}
The subsequent decay of the vector $V$ into $K\pi$ can itself be
obtained using the same approach to have the form
$K_\mu\epsilon^\mu_V$. Hence, the amplitude for $B\to \jpsi V\to
\jpsi(K \pi)_V$ may be written as\cite{Sinha:1997zu,Chiang:1999qn}:
\begin{eqnarray}
 {\cal A}_V = {\cal A} \bigg( B \to \jpsi(K \pi)_V \bigg) \propto  
   \displaystyle{\frac{ (b g_{\mu\nu} + \dsp \frac{c}{\sqrt{s_{\ell}s_K}} k_\mu q_\nu + i\dsp \frac{d}{\sqrt{s_{\ell}s_K}} \epsilon_{\mu\nu\alpha\beta} k^\alpha q^\beta )}{(k^2-M^2_V+i\epsilon)}}\,\theta^{\nu\rho}K_\rho \,\epsilon^{*\mu}_\jpsi~,
\end{eqnarray}
where,

$\Sigma\epsilon_V^{*\nu}\epsilon_V^{\rho} \equiv \theta^{\nu \rho}
=-g^{\nu \rho}+\frac{k^{\nu}k^{\rho}}{k^2}$. We clearly see that the
amplitude $B\to \jpsi V$ consists of three partial waves - $S$, $P$
and  $D$~.

Once again the amplitude for $B\to \jpsi T$ may be may written as:
\begin{eqnarray}
\label{eq:amp-T}
 {\cal A}_T &\propto&\Big(\frac{e}{\sqrt{s_{\ell} s_K}}q_\rho
 g_{\mu\nu} +\frac{f}{s_{\ell} s_K} k_\mu q_\rho q_\nu
 +i\frac{g}{s_{\ell} s_K}\epsilon_{\mu\rho\alpha\beta}q_\nu
 k^\alpha q^\beta\Big) \epsilon^{*\mu}_\jpsi
     \epsilon_T^{*\nu\rho}~,
\end{eqnarray}
where $\epsilon_T^*$ is the polarization of tensor meson. In writing 
Eq.\ref{eq:amp-T}, we used the symmetry of $\epsilon_T^{*\nu\rho}$ and
retain only terms that contribute. The amplitude for the subsequent
decay of the tensor $T$ into $K\pi$ must be of the form
$\displaystyle\epsilon^{\lambda\sigma}_T\, (k_\lambda k_\sigma
+k_\sigma K_\lambda -k_\lambda K_\sigma-K_\lambda K_\sigma)$. 
 Therefore, the amplitude for $B\to \jpsi T\to\jpsi (K
\pi)_T$ may be written as:
\begin{eqnarray}
\label{eq:AT}
 {\cal A}_T &\propto&\Big(\frac{e}{\sqrt{s_{\ell} s_K}}q_\rho
 g_{\mu\nu} +\frac{f}{s_{\ell} s_K} k_\mu q_\rho q_\nu
 +i\frac{g}{s_{\ell} s_K}\epsilon_{\mu\rho\alpha\beta}q_\nu k^\alpha q^\beta\Big)\nn\\
 &&\displaystyle{\frac{\Theta^{\nu\rho\lambda\sigma}}{k^2 - M^2_T+i\epsilon}
(k_\lambda k_\sigma +k_\sigma K_\lambda -k_\lambda K_\sigma
 -K_\lambda K_\sigma)}\epsilon^{*\mu}_\jpsi~,
\end{eqnarray}
 where
$\Sigma\epsilon_T^{*\nu\rho}\epsilon^{\lambda \sigma}_T =
\Theta^{\nu\rho\lambda\sigma}\equiv\frac{1}{2}
(\theta^{\lambda \nu}\theta^{\sigma \rho}+\theta^{\sigma
  \nu}\theta^{\lambda \rho})
-\frac{1}{3}\theta^{\lambda\sigma}\theta^{\nu
  \rho}$\cite{Spehler:1991yw}. From \ref{eq:AT}, The amplitude  $B\to\jpsi T$
 consists of three waves -$P$, $D$ and $F$ waves; The form factors ,$e$
 and $f$, are the mixture of $P$ and $F$ partial waves and the form
 factor,$g$,is related to the $CP$-odd, $D$ partial wave. It is shown in
 \ref{sec3} that all these partial waves can be extracted~.

We next incorporate the decay of $\jpsi \to \ell^+\ell^-$ which is
common to all the three decay channels. The amplitude for 
$\jpsi \to \ell^+\ell^-$is:
\begin{eqnarray}
A(\jpsi \to \ell^+\ell^-)&\propto&\epsilon^{\mu^\prime}_\jpsi(\bar{u}(q_1)\gamma_{\mu^\prime} v(q_2))~.
\end{eqnarray}
   Hence,the full decay process $B\to K(k_1)
\pi(k_2) \ell^{-}(q_1)\ell^{+}(q_2)$ may be described by the matrix
element
\begin{equation}
  {\cal M}=\frac{G_{ F}\, \alpha}{\sqrt{2}\pi}\,H_\mu
\Big(\frac{-g^{\mu\mu^\prime}+\frac{\dsp q^\mu q^{\mu^\prime}}{q^2}}{q^2-M_\jpsi^2 +i \epsilon}\Big)(\bar{u}(q_1)\gamma_{\mu^\prime} v(q_2))~,
\end{equation}
where  $H_\mu$ is the hadronic part of the matrix element and is the
sum of individual contributions from the scalar, vector and tensor
mesons. $H_\mu$ as derived above, depends on 7 independent form factors
and is given by,
\begin{eqnarray} 
H_\mu & \propto &
a k_\mu \frac{1}{(k^2-M^2_S+i\epsilon)} \nn \\
&&+\displaystyle{\frac{ (b g_{\mu\nu} + \dsp \frac{c}{\sqrt{s_{\ell}s_K}} k_\mu q_\nu + i\dsp \frac{d}{\sqrt{s_{\ell}s_K}} \epsilon_{\mu\nu\alpha\beta} k^\alpha q^\beta )}{(k^2-M^2_V+i\epsilon)}}\,\theta^{\nu\rho}K_\rho \nn\\ 
&&+\Big(\frac{e}{\sqrt{s_{\ell} s_K}}q_\rho
 g_{\mu\nu} +\frac{f}{s_{\ell} s_K} k_\mu q_\rho q_\nu
 +i\frac{g}{s_{\ell} s_K}\epsilon_{\mu\rho\alpha\beta}q_\nu k^\alpha q^\beta\Big)\nn\\
 &&\displaystyle{\frac{\Theta^{\nu\rho\lambda\sigma}}{k^2 - M^2_T+i\epsilon}
(k_\lambda k_\sigma +k_\sigma K_\lambda -k_\lambda K_\sigma
 -K_\lambda K_\sigma)}~, 
\end{eqnarray}
In what follows we assume that the mass of scalar, vector and tensor
particles are approximately equal i.e. $M_S \approx M_V \approx M_T =
M $ and define new form factors $A$, $B$, $C$, $D$, $E$, $F$, $G$ so
as to absorb the factor $\dsp \frac{1}{(k^2-M^2
  +i\epsilon)(q^2-M_\jpsi^2 +i\epsilon)}$. Defining, $H^\prime_\mu =
\dsp \frac{H_\mu}{q^2-M_\jpsi^2~.  +i\epsilon}$, we have,
\begin{eqnarray}
H^\prime_\mu &=&A k_\mu +(B g_{\mu\nu} +
\frac{C}{\sqrt{s_{\ell}s_K}} k_\mu q_\nu +
i\frac{D}{\sqrt{s_{\ell}s_K}} \epsilon_{\mu\nu\alpha\beta} k^\alpha
q^\beta )\,\theta^{\nu\rho}K_\rho\nn\\
 &&+ (\frac{E}{\sqrt{s_{\ell} s_K}}q_\rho
 g_{\mu\nu} +\frac{F}{s_{\ell} s_K} k_\mu q_\rho q_\nu
 +i\frac{G}{s_{\ell} s_K}\epsilon_{\mu\rho\alpha\beta}q_\nu k^\alpha q^\beta)\nn\\
 &&\Theta^{\nu\rho\lambda\sigma}(k_\lambda k_\sigma +k_\sigma
 K_\lambda -k_\lambda K_\sigma -K_\lambda K_\sigma)~,
\end{eqnarray}
The matrix element is finally expressed as:
\begin{eqnarray}
 \label{eq:mat-el}
  {\cal M}=\frac{G_{ F}\, \alpha}{\sqrt{2}\pi}\,H^\prime_\mu
(-g^{\mu\mu^\prime}+\frac{\dsp q^\mu q^{\mu^\prime}}{q^2})(\bar{u}(q_1)\gamma_{\mu^\prime} v(q_2))~.
\end{eqnarray}

Having obtained the matrix element for the process $B\to K(k_1)
\pi(k_2) \ell^{-}(q_1)\ell^{+}(q_2)$ it is straight forward to write
the matrix element $\overline{{\cal M}}$, for the conjugate process
$\bar{B}\to \bar{K}(k_1) \bar{\pi}(k_2) \ell^{-}(q_1)\ell^{+}(q_2)$ as
\begin{equation}
  \label{eq:barmat-el}
  \overline{\cal M}=\frac{G_{ F}\, \alpha}{\sqrt{2}\pi}\,\overline{H^\prime_\mu}( -g^{\mu\mu^\prime}+\frac{ q^\mu q^{\mu^\prime}}{q^2})(\bar{u}(q_1)\gamma_{\mu^\prime} v(q_2))~,
\end{equation}
where, $\overline{H^\prime_\mu}=\dsp \frac{\overline{H}_\mu}{(q^2-M_\jpsi^2 +i
  \epsilon)}$.   Where $\overline{H}_\mu$ is the hadronic part of the matrix element for the conjugate process and is the
sum of individual contributions from the scalar, vector and tensor
mesons. $\overline{H^\prime_\mu}$ depends on the same 7 independent
form factors and using $CPT$ may be expressed as,
\begin{eqnarray}
\overline{H^\prime_\mu} & = & A k_\mu +(B g_{\mu\nu} +
\frac{C}{\sqrt{s_{\ell}s_K}} k_\mu q_\nu -
i\frac{D}{\sqrt{s_{\ell}s_K}} \epsilon_{\mu\nu\alpha\beta} k^\alpha
q^\beta )\,\theta^{\nu\rho}K_\rho\nn\\
 &&+ (\frac{E}{\sqrt{s_{\ell} s_K}}q_\rho
 g_{\mu\nu} +\frac{F}{s_{\ell} s_K} k_\mu q_\rho q_\nu
 -i\frac{G}{s_{\ell} s_K}\epsilon_{\mu\rho\alpha\beta}q_\nu k^\alpha q^\beta)\nn\\
 &&\Theta^{\nu\rho\lambda\sigma}(k_\lambda k_\sigma +k_\sigma K_\lambda -k_\lambda K_\sigma
 -K_\lambda K_\sigma)~.
\end{eqnarray}
The time dependent study of the process $B\to K(k_1) \pi(k_2)
\ell^{-}(q_1)\ell^{+}(q_2)$ requires the knowledge of the matrix
element for both the process and the conjugate process. We can thus
study the complete time dependent angular decay of the process.
However, in this section we will derive the time integrated decay rate
and leave the time dependent study until Sec. \ref{time}.

The modulas squared of the total matrix element after summing over
lepton polarizations is
\begin{equation}
  \label{eq:mat-el-modsq}
|{\cal M}|^{2}=\frac{G_F^2 \alpha^2}{2 \pi^2}\,H^\prime_{\mu} 
  {H^{\prime\dagger}_{\mu^\prime}} \,L^{\mu\mu^\prime}~,
\end{equation}
with the leptonic tensor $L^{\mu\mu^\prime}$ being given by
\begin{equation}
L^{\mu\mu^\prime}= q^{\mu}q^{\mu^\prime}-Q^{\mu}Q^{\mu^\prime}-
     g^{\mu\mu^\prime}q^{2}~.
\end{equation}
Next we cast all the Lorentz scalars in terms of the kinematical variables.
 It is straight-forward to derive that\cite{Kruger:1999xa,Heiliger:1993qt}:
\begin{eqnarray}
  \label{eq:scalars}
  k.q &=& \frac{1}{2}(M^2_B -s_K -s_\ell)\equiv  x \sqrt{s_\ell s_K}\nn\\
  k.Q &=& X \cos\theta_{\ell}\nn\\
  q.K &=& \xi q.k +\beta X \cos\theta_K\nn\\
  K.Q &=& \xi k.Q +\beta (k.q\cos\theta_K\cos\theta_{\ell} -\sqrt{s_{\ell}s_K} \sin\theta_K \sin\theta_{\ell} \cos\phi)\nn\\
\epsilon_{\mu\nu\alpha\beta}k^{\mu}K^{\nu}q^{\alpha}Q^{\beta} & =&  -\sqrt{s_{\ell}s_{k}}\beta X\sin\theta_K\sin\theta_{\ell}\sin\phi\nn\\
 X &=& (k.q^2 -s_{\ell} s_K)^{1/2} = \frac{1}{2} \lambda^{1/2}(M_B^2,s_{\ell},s_K)\\
k.K &=& \xi s_K\nn\\
q.Q&=&0
\end{eqnarray}
where,
\begin{eqnarray}
  \label{eq:defs}
\beta&=&\frac{\lambda^{1/2}(k^2,M^2_K,M^2_\pi)}{k^2}\\
\xi&=&\frac{(M_K^{2}-M_\pi^{2})}{s_{k}}\\
\lambda(a,b,c)&=& a^2 +b^2 +c^2 -2ab -2bc -2ca 
\end{eqnarray}
and we have set $ k^2_1 = M^2_K$,~~~$ k^2_2 =M^2_\pi$.

The matrix element modulas squared obtained in Eq.~(\ref{eq:mat-el-modsq})
may finally be written in terms of the kinematical variables $s_K$,
$s_\ell$, $\theta_K$, $\theta_\ell$ and $\phi$ as follows:
\begin{eqnarray} 
\label{eq:mod.sq.}
 |{\cal M}|^{2}&=&\frac{G_F^2\alpha^2}{2
   \pi^2}( f_{0} +f_{1}\cos\phi
 +f_{2}\cos2\phi +f_{3}\sin\phi +f_{4}\sin2\phi)~, 
\end{eqnarray}
where,
\begin{equation}
  \label{eq:fp}
f_{p}=\sum(a_{mn}^{p}\cos{m \theta_{\ell}}\cos{n \theta_K}
   +b_{mn}^{p}\sin{m \theta_{\ell}}\sin{n \theta_K})~.
\end{equation}
We note that terms proportional to $\sin\theta_\ell$ are also
proportional to $\sin\theta_K$. It can be seen from 
Eq.~(\ref{eq:scalars}) that none of the Lorentz
scalars is proportional to just one of either $\sin\theta_\ell$ or
$\sin\theta_K$.  Hence there can be no term odd in $\theta_\ell$ and
even in $\theta_K$ or vice versa -- only terms even in both
$\theta_\ell$ and $\theta_K$ or odd in both $\theta_\ell$ and
$\theta_K$ are possible. The form chosen for $f_p$ in
Eq.~(\ref{eq:fp}) is thus easily understood. All the non vanishing
co-efficients $a_{mn}^p$ and $b_{mn}^p$ are listed under
Tables~\ref{table1}-\ref{table6} in the Appendix. Identical results
were obtained using the helicity formalism \cite{BaBar-collaboration}~.

\section{Solution of the form factors and Analysis}
\label{sec3}
It was seen in Sec. \ref{sec2} that angular analysis can be used to
obtain coefficients $a^p_{mn}$ and $b^p_{mn}$. We begin this section
by demonstrating the explicit solutions that 
 enable us to obtain both the magnitudes and phases of the
form factors $a$, $b$, $c$, $d$, $e$, $f$ and $g$. 
 We show that angular analysis for this mode  will 
enable us to isolate the contributions from each of the partial-waves.
We can thus separate the contributions of the partial waves into
$CP$ even and $CP$ odd, allowing us to measure the mixing phase
$\beta$ without worrying about dilution from the wrong $CP$ parity
contributions. In this section we also consider both the time integrated
differential decay rate and time dependent differential decay rate. We
study various angular asymmetries and point out certain interesting
features that the angular study should obey. We also construct a time
dependent angular asymmetry that enables a clean measurement of the
mixing phase $\beta$ in the mode $B\to \jpsi \Kstart(1430)$ alone
without contributions from the decay modes $B\to \jpsi\Kstarsz(1430)$
and $B\to\jpsi\Kstar(1410)$.
\subsection{Solution of the form factors\label{solutions-ff}}

We note that the angular analysis allows one to measure a large number
of coefficients $a^p_{mn}$, $b^p_{mn}$. In spite of several of these
coefficient, being zero, there is still enough information to solve
for both the magnitude and phase of all the form factors.  In fact,
the zero values act as constraints and are hence useful in
experimental fits to the coefficients.  We do not explicitly integrate
over $s_\ell$ and $s_K$ in our discussions, however, for the purpose
of experimental analysis these variables may be integrated over,
before extracting the form factors. In this subsection we present
explicit solutions to the magnitudes and phases of the form factors,
and thereby established that an angular analysis allows us to
disentangle the contribution from each of the three resonances
considered.  To set up our notation we define $A=|A|\exp (i\phi_A)$,
$B=|B|\exp (i\phi_B)$, $C=|C|\exp (i\phi_C)$, $D=|D|\exp (i\phi_D)$,
$E=|E|\exp (i\phi_E)$, $F=|F|\exp (i\phi_F)$ and $G=|G|\exp
(i\phi_G)$. Using the tables \ref{table1}-\ref{table6}, it is
straightforward to verify that
$|B|$,~$|D|$,~$|E|$,~$|F|$,~$|G|$
are given in terms of experimentally obtainable coefficients by,
\begin{eqnarray}
|B|^2 & =
&-\frac{2}{\,\beta^{2}s_{\ell}\,s_{K}}(a^{0}_{22}+a^{0}_{02}-4a^{2}_{02})\\
|D|^2 & =
&-\frac{2}{\,\beta^{2}s_{\ell}\,s_{K}(x^2-1)}(a^{0}_{22}+a^{0}_{02}+4a^{2}_{02})\\
|E|^2& = &\frac{32}{\,\beta^{4}s_{\ell}\,s_{K}(x^2-1)}
(4a^{2}_{04}-(a^{0}_{04}+a^{0}_{24}))\\
|F|^2& =
&\frac{32}{\,\beta^{4}s_{\ell}\,s_{K}(x^2-1)^3}(4x^2\,a^{2}_{04}-(x^2+3)\,a^{0}_{24}-(x^2-1)a^{0}_{04}-4x\,b^{1}_{24})\\
|G|^2& = &\frac{-32}{\,\beta^{4}s_{\ell}\,s_{K}(x^2-1)^2}
(4a^{2}_{04}+a^{0}_{04}+a^{0}_{24})
\end{eqnarray}
Also, $\Im(BD^{\star})$, $\Re(BE^{\star})$, $\Re(DG^{\star})$,
$\Im(EG^{\star})$, $\Re(FE^{\star})$ and $\Im(FG^{\star})$, can easily
be written in terms of observables as:
\begin{eqnarray}
\Im(EG^{\star})& = &\frac{128}{\,\beta^{4}s_{\ell}\,s_{K}(x^2-1)^{3/2}}\,a^{4}_{04}\\
\Im (FG^{\star}) & =
&\frac{64}{\,\beta^{4}s_{\ell}\,s_{K}(x^2-1)^{5/2}}(b^{3}_{24}-2x\,a^{4}_{04})\\
\Im(BD^{\star})& =
&\frac{8}{\,\beta^{2}s_{\ell}\,s_{K}\sqrt{(x^2-1)}}\,a^{4}_{02}\\
\Re(FE^{\star}) & =
&\frac{32}{\,\beta^{4}s_{\ell}\,s_{K}(x^2-1)^2}(2b^{1}_{24}+x(a^{0}_{04}+a^{0}_{24})-4x\,a^{2}_{04})\\
\label{eq:CD}
\Re(CD^{\star})& =
&\frac{-4}{\,\beta^{3}s_{\ell}\,s_{K}(x^2-1)^{5/2}}((x^2-1)a^{0}_{03}+(x^2+3)a^{0}_{23}+4x^2a^{2}_{23}+4xb^{1}_{23})\\
\Re(BE^{\star}) &=
&\frac{4}{\,\beta^{3}s_{\ell}\,s_{K}(x^2-1)^{1/2}}(a^{0}_{01}+a^{0}_{21}-4a^{2}_{01})\\
\Re(DG^{\star}) &=
&\frac{4}{\,\beta^{3}s_{\ell}\,s_{K}(x^2-1)^{3/2}}(a^{0}_{01}+a^{0}_{21}+4a^{2}_{01})
\end{eqnarray}
We set $\phi_G=0$ by convention. The phases
~$\phi_B$,~$\phi_D$,~$\phi_E$,~$\phi_F$ can be obtained using
$\Im(BD^{\star})$, $\Re(BE^{\star})$, $\Re(DG^{\star})$,
$\Im(EG^{\star})$, $\Re(FE^{\star})$ and $\Im(FG^{\star})$, which have
already been evaluated with respect to observables using,
\begin{eqnarray}
\phi_F -\phi_E & = &\cos^{-1}\Big(\frac{\Re(FE^{\star})}{|E|
  |F|}\Big)\\
\phi_E -\phi_G & = &\sin^{-1}\Big(\frac{\Im(EG^{\star})}{|E|
  |G|}\Big)\\
\phi_F -\phi_G & = &\sin^{-1}\Big(\frac{\Im(FG^{\star})}{|F|
  |G|}\Big)\\
\phi_B -\phi_E & = &\cos^{-1}\Big(\frac{\Re(BE^{\star})}{|B|
  |E|}\Big)\\
\phi_D -\phi_G & = &\cos^{-1}\Big(\frac{\Re(DG^{\star})}{|D|
  |G|}\Big)\\
\phi_B -\phi_D & = &\sin^{-1}\Big(\frac{\Im(BD^{\star})}{|B|
  |D|}\Big)
\end{eqnarray}
Note that ambiguities in the solution are somewhat reduced due to the
constraints already obtained in the above equations.  The resolution
of ambiguities is not discussed in detail here. However, it may be
envisaged that the approach used by the BaBar collaboration
\cite{BaBar-collaboration} can also be applied here to completely
remove ambiguities in the present study.  The remaining form factors
can also be solved as follows:
\begin{eqnarray}
\label{eq:CE}
\Re(CE^{\star})& =&
\frac{1}{2 s_{\ell}\,s_{K}\,
  x(x^2-1)^{3/2}}\Big(-32a^{0}_{23}-\sqrt{x^2-1}\, \beta^3(2\, x
  (x^2-1) \Re(BF^\star)+\nn \\
& &(2x^2+1)\Re(BE^\star)+ (x^2-1)(2(x^2-1)
  \Re(CD^\star)+\Re(DG^\star)))s_{\ell}s_K \Big)
\end{eqnarray}
Having obtained the value of
$|B|$,~$|D|$,~$|E|$,~$|F|$,~$|G|$,~$\phi_B$,~$\phi_D$,~$\phi_E$,~$\phi_F$
and ~$\phi_G$, ~$\Re(BE^\star) =|B| |E| \cos(\phi_B -\phi_E)$ and
$\Re(BF^\star)=|B| |F| \cos(\phi_B-\phi_F)$ are in fact, also obtained
in terms of observables.  On putting these values in
Eq.~(\ref{eq:CE}), $\Re(CE^{\star})$ is evaluated in terms of
observables. $\phi_C$ and $|C|$ can now be solved for using
Eqs.~(\ref{eq:CD}) and (\ref{eq:CE}), to give,
 \begin{eqnarray}
\phi_C & = &\sin^{-1}\Big(\dsp \frac{\cos\phi_E
  -\eta\cos\phi_D}{\sqrt{1+\eta2 -2\eta
    \cos(\phi_D-\phi_E)}}\Big)\\
|C| & = & \frac{\Re(CD^\star)}{|D| \cos(\phi_C-\phi_D)} 
\end{eqnarray}
where,
\[
\eta = \dsp \frac{|D|\Re(CE^\star)}{|E|\Re(CD^\star)}
\]
Using a similar procedure one can also solve for the only remaining
quantities $\phi_A$ and $|A|$. We have,
\begin{eqnarray}
\label{eq:AE}
\Re(AE^{\star})
&=&\frac{1}{12\,\beta^{2}s_{\ell}\,s_{K}(x^2-1)}\Big(-96 \,b^1_{22}
  +\beta^{2}s_{\ell}\,s_{K}(24 |B|^2 +\beta^{2}(x^2 -1)|E|^2+\nn \\
& & 24(x^2-1)\Re(BC^\star)+\beta^{2}(x^2 -1)^2 \, \Re(FE^\star))
\Big)\\
\label{eq:AB}
\Re(AB^{\star})
&=&\frac{1}{12\,\beta s_{\ell}\,s_{K}\sqrt{x^2-1}} \Big(-48\,
b^1_{21}
+\beta^{3}\sqrt{x^2-1}\,s_{\ell}\,s_{K}\nn \\
& &((x^2-1)(3\Re(CE^{\star})
-\Re(BF^{\star})) +2 x \,\Re(BE^{\star}))\Big)
\end{eqnarray}
With the values of $\phi_C$ and $|C|$ already obtained,
$\Re(BC^\star)$ and $\Re(CE^{\star})$ are also expressed in terms of
observables. We finally have after solving Eqs.~(\ref{eq:AE}) and
(\ref{eq:AB}):
\begin{eqnarray}
\phi_A & = &\sin^{-1}\Big(\frac{\cos\phi_E
  -\varsigma\cos\phi_B}{\sqrt{1+\varsigma^2 -2\varsigma
    \cos(\phi_B-\phi_E)}}\Big)\\
|A| & = & \frac{\Re(AB^\star)}{|B| \cos(\phi_A-\phi_B)} 
\end{eqnarray}
where,
\[
\varsigma = \dsp \frac{|B|\Re(AE^\star)}{|E|\Re(AB^\star)}
\]

\subsection{Time Integrated Differential Decay Rate}

The differential decay rate is\\
\begin{equation}
 d\Gamma = \frac{\beta X }{2^{15} \pi^6 M_B^3} |{\cal M}|^{2} ds_{\ell}~ ds_K~ d\cos\theta_{\ell} ~ d\cos\theta_K ~ d\phi~.
\end{equation}
Therefore,
\begin{eqnarray}
 \displaystyle{\frac{d\Gamma}{ds_{\ell}~ ds_K~d\cos\theta_{\ell} ~ d\cos\theta_K ~ d\phi}} ~~~~~= \frac{\beta X }{2^{15} \pi^6 M_B^3} |{\cal M}|^{2}
\end{eqnarray}
The physical regions of the angular variables are 
\begin{equation}
0 \leq \phi \leq 2\pi,~~~-1 \leq \cos\theta_K \leq 1 ~~~\mathrm{and}~~~-1 \leq \cos\theta_{\ell} \leq 1~,
\end{equation}
and $s_\ell$ and $s_K$ are integrated over the relevant resonances.
We derive the one-dimensional angular distributions
$\displaystyle{d\Gamma/(ds_{\ell}~ ds_K~ d\cos\theta_{\ell})}$,
$\displaystyle{d\Gamma/(ds_{\ell}~ ds_K~d\cos\theta_K)}$, and
$\displaystyle{d\Gamma/(ds_{\ell}~ ds_K~d\phi)}$ from the differential
decay rate. These distributions as well as the other observables
extracted by the angular analysis, depend on different combinations of
the co-efficients $a^p_{mn}$and $b^p_{mn}$.

\subsubsection{Decay rate as a function of $\cos\theta_K$}
Integrating the Eq.~(\ref{eq:mod.sq.}) over $\cos\theta_{\ell}$ and $\phi$  we obtain 
\begin{eqnarray}
\displaystyle{\frac{d\Gamma}{ds_{\ell}~ds_K~d\cos\theta_K}}&=&\frac{\beta
  X G_F^2 \alpha^2 }{2^{16} \pi^8 M_B^3} \frac{4 \pi}{3}\{3a^{0}_{00} -a^{0}_{20}+(3a^{0}_{01}-a^{0}_{21})\cos\theta_K +(3a^{0}_{02}-a^{0}_{22})\cos2\theta_K\nonumber\\
&& +(3a^{0}_{03}-a^{0}_{23})\cos3\theta_K +(3a^{0}_{04}-a^{0}_{24})\cos4\theta_K\} 
\end{eqnarray}
Now we define the forward--backward(FB) asymmetry in $ K\pi$ system
\begin{eqnarray}
A^K_{FB} &=&\frac{\int_0^1\displaystyle{\frac{d\Gamma}{ds_{\ell}~ds_K~d\cos\theta_K}}d\cos\theta_K -\int_{-1}^0\displaystyle{\frac{d\Gamma}{ds_{\ell}~ds_K~d\cos\theta_K}}d\cos\theta_K}{\int_0^1\displaystyle{\frac{d\Gamma}{ds_{\ell}~ds_K~d\cos\theta_K}}d\cos\theta_K +\int_{-1}^0\displaystyle{\frac{d\Gamma}{ds_{\ell}~ds_K~d\cos\theta_K}}d\cos\theta_K}\nonumber\\
&=&\frac{15}{2}\frac{3(a^{0}_{01}-a^{0}_{03})+a^{0}_{23}-a^{0}_{21}}{45a^{0}_{00}-15(a^{0}_{02}+a^{0}_{20})-3a^{0}_{04}+5a^{0}_{22}+a^{0}_{24}}\nonumber\\
&\neq&0
\end{eqnarray}
This is not vanishing due to the presence of
$\cos\theta_K$ and $\cos3\theta_K$. These terms are present due to interference between
`vector and scalar' as well as between `vector and tensor' mesons
contributions as intermediate states. However, the forward--backward
asymmetry vanishes in each of the $B\to\jpsi S$, $B\to \jpsi V$ and
$B\to \jpsi T$ decay modes.
\subsubsection{Decay rate as a function of $\cos\theta_{\ell}$}
Integrating the Eq.~(\ref{eq:mod.sq.}) over $\cos\theta_K$ and $\phi$ we obtain
\begin{eqnarray}
\frac{d\Gamma}{ds_{\ell}~ds_K~d\cos\theta_{\ell}} &=&\frac{\beta X
  G_F^2 \alpha^2 }{2^{16} \pi^8 M_B^3 }
\frac{4\pi}{15}\{15a^{0}_{00}-5a^{0}_{02}-a^{0}_{04}\nonumber\\
&&+(15a^{0}_{20}-5a^{0}_{22}-a^{0}_{24})\cos2\theta_{\ell}\}
\end{eqnarray}
It is easy to see that the forward-backward(FB) asymmetry in the
$\ell^- \ell^+$ system vanishes, i.e. $A^{\ell}_{FB}=0$. 
The absence term odd in $\cos\theta_{\ell}$ is connected with the
fact that the $\ell^- \ell^+$ system is in a pure $L=1$ state. As a
consequence, the forward--backward(FB) asymmetry in the system vanishes.
\subsubsection{Decay rate as a function of $\phi$}
Finally, the distribution in the angle $\phi$ between the lepton and
meson planes, after integration of the Eq.~(\ref{eq:mod.sq.})  over
$\cos\theta_{\ell}$ and $\cos\theta_k$, we obtain
\begin{eqnarray}
\frac{d\Gamma}{ds_{\ell}~ds_K~d\phi}&=&\frac{\beta  X G_F^2 \alpha^2 }{2^{16} \pi^8 M_B^3} \frac{4}{45}\{45a^{0}_{00}-15(a^{0}_{02}+a^{0}_{20})+5a^{0}_{22}-3a^{0}_{04}+a^{0}_{24} \nonumber\\
& &+(45a^{2}_{00}-15(a^{2}_{02}+a^{2}_{20})+5a^{2}_{22}-3a^{2}_{04}+a^{2}_{24})\cos2\phi\nonumber\\
& &+(45a^{4}_{00}-15(a^{4}_{02}+a^{4}_{20})+5a^{4}_{22}-3a^{4}_{04}+a^{4}_{24})\sin2\phi\}
\end{eqnarray}
The presence of $\sin2\phi$ term is a clean signal of CP violation in
the $\phi$ distribution in the decay process.
\subsection{Time Dependent Differential decay rate and measurement of $\boldmath \beta$}\label{time}

In this subsection we demonstrate how a clean measurement of $\sin
2\beta$ can be performed using one of the partial wave contributing to
$B\to\jpsi\Kstart(1430)$. For this purpose we use the $CP$-even
partial waves contributing to $B\to\jpsi\Kstart(1430)$. In
Sec.~\ref{solutions-ff} we showed $|G|^2$ can be extracted using the
measurements of coefficients $a^{0}_{04}$, $a^{0}_{24}$, $a^{2}_{04}$.
Here, we derive an angular and time-dependent asymmetry that cleanly 
measures $\sin
2\beta$, without the need for a complete solutions to all the
contributing form factors.
It is straight
forward to see that the coefficient of time dependent $\sin(\Delta
M\,t)$ term corresponding to $a^0_{04}$, $a^0_{24}$ and $a^2_{04}$
 of Table \ref{table1}, \ref{table2} and \ref{table3} can be obtained
 by the replacement $|G|^2\to|G|^2\sin 2\beta$.  A time dependent angular
analysis can be performed to isolate the $\sin(\Delta M\,t)$ term to
$a^{0}_{04}$, $a^{0}_{24}$, $a^{2}_{04}$. An asymmetry that isolates
 such a term is given by:
\begin{eqnarray}
 \label{eq:acp}
A_{CP}\,\sin(\Delta M\,t) & = & \frac{1}{\dsp\int 
  _{-1}^{1}d\cos\theta_K \int _{-1}^{1}
  d\cos\theta_\ell\int_{0}^{2\pi}
    d\phi\,\Big(\frac{d(\Gamma(t)+\bar{\Gamma}(t))}
    {d\cos\theta_\ell\,d\cos\theta_K\,d\phi}\Big)}\,\times \nn\\ 
& &\Big(\int _P\dsp
  d\cos\theta_K \int _{-1}^{1} d\cos\theta_\ell \int _{0}^{2\pi} d\phi+
  \int _P d\cos\theta_K \int _T d\cos\theta_\ell \int _{0}^{2\pi}d\phi+\nn\\ 
& &
\int _P\dsp d\cos\theta_K \int _{-1}^{1} d\cos\theta_\ell \int _Q d\phi\Big)
\Big(\dsp\frac{d(\Gamma(t)-\bar{\Gamma}(t))}
  {d\cos\theta_\ell\,d\cos\theta_K\,d\phi}\Big)~.
\end{eqnarray}
where, 
\begin{eqnarray}
  \label{eq:integrals}
  \int _P d\cos\theta_K &=&(\int _{0}^{\frac{\pi }{5}}- 
\int _{\frac{\pi }{5}}^{\frac{2\,\pi }{5}}+ 
\int _{\frac{2\,\pi }{5}}^{\frac{3\pi }{5}}- 
\int _{\frac{3\pi }{5}}^{\frac{4\,\pi }{5}}+\int _{\frac{4\,\pi }{5}}^{\pi })(-\sin\theta_K )d\theta_K\\
\int _{T} d\cos\theta_\ell &=& (\int _{0}^{\frac{\pi}{3}} -\int _{\frac{\pi}{3}}^{\frac{2\,\pi}{3}}+\int _{\frac{2\pi}{3}}^{\pi})(-\sin\theta_{\ell} )d\theta_\ell,\\
\int _Q d\phi& =&(\int _{\frac{\pi}{4}}^{\frac{3\pi }{4}}-\int _{\frac{3\pi }{4}}^\frac{5\pi }{4}{} +\int _{\frac{5\pi }{4}}^{\frac{7\pi }{4}} -\int _{\frac{7\pi }{4}}^{\frac{9\pi}{4}})d\phi~.
\end{eqnarray}
In Eq.~(\ref{eq:acp}) integrals over the relevent range for $s_\ell$
and $s_K$ are implicit.  Using Eqns.~(\ref{eq:mod.sq.}) and
(\ref{eq:fp}) Table \ref{table1}-\ref{table5}, we see that
\begin{eqnarray}
  \label{eq:theo}
  A_{CP}={\cal R}\,\sin 2\beta~,
\end{eqnarray}
where
\begin{eqnarray}
{\cal R} & \propto & \dsp \frac{4\,a^2_{04}
  +a^0_{04}+a^0_{24}}{45a^{0}_{00}-15(a^{0}_{02}+a^{0}_{20})-3a^{0}_{04}+5a^{0}_{22}+a^{0}_{24}}~.
\end{eqnarray}
$\cal R$ can itself be obtained directly from experimental data by an
analogous asymmetry using time integrated partial decay rates as
follows:
\begin{eqnarray}
  \label{eq:R2}
  {\cal R}& = &\frac{1}{\dsp\int 
  _{-1}^{1}d\cos\theta_K \int _{-1}^{1}
  d\cos\theta_\ell\int_{0}^{2\pi}
    d\phi\,\Big(\frac{d\Gamma}
    {d\cos\theta_\ell\,d\cos\theta_K\,d\phi}\Big) }\,\times\nn\\ 
& &\Big(\int _P\dsp
  d\cos\theta_K \int _{-1}^{1} d\cos\theta_\ell \int _{0}^{2\pi} d\phi+
  \int _P d\cos\theta_K \int _T d\cos\theta_\ell \int _{0}^{2\pi}d\phi+\nn\\ 
& &
\int _P\dsp d\cos\theta_K \int _{-1}^{1} d\cos\theta_\ell \int _Q d\phi\Big)
\Big(\dsp\frac{d\Gamma}
  {d\cos\theta_\ell\,d\cos\theta_K\,d\phi}\Big)~.   
\end{eqnarray}
We can thus obtain a clean measurement of $\sin(2\beta)$ using the
$CP$--even part of the mode $B\to \jpsi \Kstart(1430)$ alone, without
any contributions from the decay modes $B\to\jpsi\Kstarsz(1430)$ or
$B\to\jpsi\Kstar(1410)$.

\section{Conclusion}
\label{sec4}

We have studied how an angular analysis can be used to isolate
contributions from each of the decay modes $B\to \jpsi\Kstarsz(1430)$,
$B\to\jpsi\Kstar(1410)$ and $B\to \jpsi \Kstart(1430)$, where
$\Kstarsz(1430)$, $\Kstar(1410)$ and $\Kstart(1430)$ are overlapping
and contribute to the same final state $B\to K\pi\ell^+\ell^-$.
Angular analysis also allows us to isolate the contributions from
different partial waves contributing to the each of these final
states. We have studied the time integrated differential decay rate
and derived explicit solutions to both the magnitudes and the phases of
the form factors contributing.  We showed that the
forward-backward(FB) asymmetry in $ \ell^+ \ell^-$ system vanishes,
since the $\ell^- \ell^+$ system is in a pure $L=1$ state. We have
also studied the forward-backward(FB) asymmetry in $K \pi$ system;
such terms are present due to interference between contributions from
vector and scalar, as well as between vector and tensor intermediate
states. We also construct a time dependent angular asymmetry that
enables a clean measurement of the mixing phase $\beta$ in the mode
$B\to \jpsi \Kstart(1430)$ alone, without contributions from the decay
modes $B\to\jpsi\Kstarsz(1430)$ or $B\to\jpsi\Kstar(1410)$.
The study performed here finds immediate application in
the analysis of data collected by the Belle and BaBar collaborations.

\section{Acknowledgments}
It is a pleasure to thank W. Dunwoodie, J. B. Singh and Nitesh Soni
for discussions which motivated us to look at this problem. We also
thank W. Dunwoodie for several other suggestions.

\vspace{2cm}
\newpage
\section{Appendix}

\begin{table}[htbp]
  \centering
  \begin{tabular}{| c |lr|}    
    \hline
&&\\
~~$\dsp a^{0}_{00}$~~
& $\dsp\frac{1}{2304}\,s_{\ell}\,s_{K}\Big((576|A|^2(x^{2}-1)
        +144\,(|B|^2(2x^{2}+3)+(x^{2}-1)\,(3|D|^2+2|C|^2(x^{2}-1)))\beta^{2}$
    &\\& $
 +(x^{2}-1)\,(|E|^{2}(27+22x^2)+(x^{2}-1)(27\,|G|^{2}+22\,|F|^{2}(x^{2}-1)))\beta^{4}$
    &\\& $+4(x^{2}-1)\beta^{2}(-24(x^{2}-1)\,\Re(A\,F^{\star})+x(-24\,\Re(A\,E^{\star})+144\,\Re(B\,C^{\star})$
    &\\&~~~~~~~~~~~~~~~~~~~~~~~~~~~~~~~~ $+11(x^{2}-1)\beta^{2}\,\Re(F\,E^{\star})))\Big) $
&
\\  

&&\\
~~$\dsp a^{0}_{01}$~~
&  $ \dsp\frac{1}{96}\,\beta\sqrt{x^{2}-1}\,s_{\ell}\,s_{K}\Big(-48x\,\Re(A\,B^{\star})-48(x^{2}-1)\,\Re(A\,C^{\star})+\beta^{2}(10x(x^{2}-1)\,\Re(B\,F^{\star})$
    &\\& $+(10x^2+9)\,\Re(B\,E^{\star})+(x^{2}-1)(10((x^{2}-1)\,\Re(C\,D^{\star})+10x\,\Re(C\,E^{\star})+9\,\Re(D\,G^{\star})))\Big)$ 
&\\

&&\\
~~$\dsp a^{0}_{02}$~~
&  $\dsp\frac{1}{96}\,\beta^{2}s_{\ell}\,s_{K}\Big(6|B|^{2}(2x^2-3)+(x^{2}-1)(-18|D|^{2}+12|C|^{2}(x^2-1)$
    &\\& $+(|E|^{2}x^2+|F|^{2}(x^2-1)^2)\,\beta^{2})-12(x^2-1)^2\,\Re(A\,F^{\star})+2x(x^2-1)(-6\,\Re(A\,E^{\star}) $
    &\\&$ +12\,\Re(B\,C^{\star})+(x^2-1)\beta^{2}\,\Re(F\,E^{\star}))\Big)$ 
&\\
 
&&\\
~~$\dsp a^{0}_{03}$~~
&  $ \dsp\frac{1}{32}\,\beta^3 s_{\ell}\,s_{K}\sqrt{x^{2}-1}\Big(2x(x^2-1)\,\Re(B\,F^{\star})+(2x^2-3)\,\Re(B\,E^{\star})$
&\\& $+(x^{2}-1)(2(x^{2}-1)\,\Re(C\,D^{\star})+2x\,\Re(C\,E^{\star})-3\Re(D\,G^{\star}))\Big)$     
&\\

&&\\
~~$\dsp a^{0}_{04}$~~
&  $\dsp\frac{1}{256}\,\beta^{4}s_{\ell}\,s_{K}(x^2-1)\Big(|E|^{2}(2x^2-3)+(x^{2}-1)(-3|G|^{2}+2|F|^{2}(x^2-1))$
    &\\& $+4x(x^2-1))\,\Re(F\,E^{\star})\Big)$ 
&\\\hline
\end{tabular}
\caption{ Non-vanishing $a^0_{0n}$} 
\label{table1}
\end{table}

 \begin{table}[htbp]
  \centering
  \begin{tabular}{| c |lr|}    
    \hline
&&\\
~~$\dsp a^{0}_{20}$~~
&  $\dsp\frac{1}{2304}\,s_{\ell}\,s_{K}\Big((-576|A|^2(x^{2}-1)
        -144\,(|B|^2(2x^{2}-1)+(x^{2}-1)\,(-|D|^2+2|C|^2(x^{2}-1)))\beta^{2}$
    &\\& $
 -(x^{2}-1)\,(|E|^{2}(-9+22x^2)+(x^{2}-1)(-9\,|G|^{2}+22\,|F|^{2}(x^{2}-1)))\beta^{4}$
    &\\& $+4(x^{2}-1)\beta^{2}(24(x^{2}-1)\,\Re(A\,F^{\star})+24x(\,\Re(A\,E^{\star})-6\,\Re(B\,C^{\star}))$
    &\\&~~~~~~~~~~~~~~~~~~~~~~~~~~~~~~~~ $-11x(x^{2}-1)\beta^{2}\,\Re(F\,E^{\star}))\Big) $ 
&\\ 

&&\\
~~$\dsp a^{0}_{21}$~~
&  $ \dsp\frac{1}{96}\,\beta\sqrt{x^{2}-1}\,s_{\ell}\,s_{K}\Big(48x\,\Re(A\,B^{\star})+48(x^{2}-1)\,\Re(A\,C^{\star})+\beta^{2}(-10x(x^{2}-1)\,\Re(B\,F^{\star})$
    &\\& $+(-10x^2+3)\,\Re(B\,E^{\star})-(x^{2}-1)(10((x^{2}-1)\,\Re(C\,D^{\star})+10x\,\Re(C\,E^{\star})-3\,\Re(D\,G^{\star})))\Big)$
&\\

&&\\
~~$\dsp a^{0}_{22}$~~
&  $-\dsp\frac{1}{96}\,\beta^{2}s_{\ell}\,s_{K}\Big(6|B|^{2}(2x^2+1)+(x^{2}-1)(6|D|^{2}+12|C|^{2}(x^2-1)$
    &\\& $+(|E|^{2}x^2+|F|^{2}(x^2-1)^2)\,\beta^{2})-12(x^2-1)^2\,\Re(A\,F^{\star})+2x(x^2-1)(-6\,\Re(A\,E^{\star}) $
    &\\&$ +12\,\Re(B\,C^{\star})+(x^2-1)\beta^{2}\,\Re(F\,E^{\star}))\Big)$ 
&\\ [1ex]

&&\\
~~$\dsp a^{0}_{23}$~~
&  $ -\dsp\frac{1}{32}\,\beta^3 s_{\ell}\,s_{K}\sqrt{x^{2}-1}\Big(2x(x^2-1)\,\Re(B\,F^{\star})+(2x^2+1)\,\Re(B\,E^{\star})$
&\\& $+(x^{2}-1)(2(x^{2}-1)\,\Re(C\,D^{\star})+2x\,\Re(C\,E^{\star})+\Re(D\,G^{\star}))\Big)$     
&\\

&&\\
~~$a^{0}_{24}$~~
&  $-\dsp\frac{1}{256}\,\beta^{4}s_{\ell}\,s_{K}(x^2-1)\Big(|E|^{2}(2x^2+1)+(x^{2}-1)(|G|^{2}+2|F|^{2}(x^2-1))$
    &\\& $+4x(x^2-1))\,\Re(F\,E^{\star})\Big)$ 
& \\ \hline
\end{tabular}
\caption{ Non-vanishing $a^0_{2n}$} 
\label{table2}
\end{table}

\begin{center}
\begin{table}[h]

\begin{tabular}{|l |lr|}   
\hline
&&\\
~~$\dsp a^{2}_{00}$~~
& $\dsp\frac{1}{256}\,\beta^{2}s_{\ell}\,s_{K}(-16|B|^2+(x^2-1)(16|D|^2-(|E|^2-|G|^2(x^2-1))\,\beta^{2})) $
 &\\  

&&\\
~~$\dsp a^{2}_{01}$~~
& $\dsp\frac{1}{32}\,\beta^{3}s_{\ell}\,s_{K} \sqrt{(x^2-1)}(-\,\Re(BE^{\star})+(x^2-1)\,\Re(DG^{\star}))$
& \\
 
&&\\   
~~$\dsp a^{2}_{02}$
& $\dsp\frac{1}{16}\,\beta^{2}s_{\ell}\,s_{K}(|B|^2-(x^2-1)|D|^2) $
& \\

&&\\  
~~$\dsp a^{2}_{03}$~~
&$-\dsp\frac{1}{32}\,\beta^{3}s_{\ell}\,s_{K} \sqrt{(x^2-1)}(-\,\Re(BE^{\star})+(x^2-1)\,\Re(DG^{\star})) $
&\\

&&\\  
~~$\dsp a^{2}_{04}$~~
& $\dsp\frac{1}{256}\,\beta^{4}s_{\ell}\,s_{K} (x^2-1)(|E|^2-|G|^2(x^2-1))$
& \\

&&\\ 
~~$\dsp a^{2}_{20}$~~
& $-\dsp\frac{1}{256}\,\beta^{2}s_{\ell}\,s_{K}(-16|B|^2+(x^2-1)(16|D|^2-(|E|^2-|G|^2(x^2-1))\,\beta^{2})) $
&\\

&&\\    
~~$\dsp a^{2}_{21}$~~
& $-\dsp\frac{1}{32}\,\beta^{3}s_{\ell}\,s_{K} \sqrt{(x^2-1)}(-\,\Re(BE^{\star})+(x^2-1)\,\Re(DG^{\star})) $
&\\

&&\\  
~~$\dsp a^{2}_{22}$~~
& $-\dsp\frac{1}{16}\,\beta^{2}s_{\ell}\,s_{K}(|B|^2-(x^2-1)|D|^2) $
& \\

&&\\   
~~$\dsp a^{2}_{23}$~~
& $\dsp\frac{1}{32}\,\beta^{3}s_{\ell}\,s_{K} \sqrt{(x^2-1)}(-\,\Re(BE^{\star})+(x^2-1)\,\Re(DG^{\star})) $
&
\\ 
&&\\   
~~$\dsp a^{2}_{24}$~~
& $-\dsp\frac{1}{256}\,\beta^{4}s_{\ell}\,s_{K} (x^2-1)(|E|^2-|G|^2(x^2-1)) $
&\\ [2ex] \hline
\end{tabular}
\caption{ Non-vanishing $a^2_{mn}$} 
\label{table3}
\end{table}
\end{center}

\begin{center}
\begin{table}[h]
\label{table4}

\begin{tabular}{|l |lr|}     
\hline
&&\\
~~$\dsp a^{4}_{00}$~~
& $-\dsp\frac{1}{128}\,\beta^{2}s_{\ell}\,s_{K} \sqrt{(x^2-1)}\Big(16\,\Im(BD^{\star})+(x^2-1)\,\beta^{2}\,\Im(EG^{\star})\Big) $
&\\

&&\\     
~~$\dsp a^{4}_{01}$~~
& $-\dsp\frac{1}{32}\,\beta^{3}s_{\ell}\,s_{K} (x^2-1)\Big(\,\Im(BG^{\star})-\,\Im(DE^{\star})\Big) $
&\\

&&\\
~~$\dsp a^{4}_{02}$~~
& $\dsp\frac{1}{8}\,\beta^{2}s_{\ell}\,s_{K} \sqrt{(x^2-1)}\,\Im(BD^{\star}) $
&\\

&&\\
~~$\dsp a^{4}_{03}$~~
& $\dsp\frac{1}{32}\,\beta^{3}s_{\ell}\,s_{K} (x^2-1)\Big(\,\Im(BG^{\star})-\,\Im(DE^{\star})\Big) $
&\\

&&\\   
~~$\dsp a^{4}_{04}$~~
& $\dsp\frac{1}{128}\,\beta^{4}s_{\ell}\,s_{K} (x^2-1)^{3/2}\,\Im(EG^{\star})$
&\\  [2ex]

&&\\ 
~~$\dsp a^{4}_{20}$~~
&$\dsp\frac{1}{128}\,\beta^{2}s_{\ell}\,s_{K} \sqrt{(x^2-1)}\Big(16\,\Im(BD^{\star})+(x^2-1)\,\beta^{2}\,\Im(EG^{\star})\Big) $
&\\

&&\\  
~~$\dsp a^{4}_{21}$~~
& $\dsp\frac{1}{32}\,\beta^{3}s_{\ell}\,s_{K} (x^2-1)\Big(\,\Im(BG^{\star})-\,\Im(DE^{\star})\Big) $
&
\\  
&&\\
~~$\dsp a^{4}_{22}$~~
& $-\dsp\frac{1}{8}\,\beta^{2}s_{\ell}\,s_{K} \sqrt{(x^2-1)}\,\Im(BD^{\star})$
&
\\  
&&\\
~~$\dsp a^{4}_{23}$~~
& $-\dsp\frac{1}{32}\,\beta^{3}s_{\ell}\,s_{K} (x^2-1)\Big(\,\Im(BG^{\star})-\,\Im(DE^{\star})\Big)$
&
\\
&&\\  
~~$\dsp a^{4}_{24}$~~
& $-\dsp\frac{1}{128}\,\beta^{4}s_{\ell}\,s_{K} (x^2-1)^{3/2}\,\Im(EG^{\star}) $
&\\ [2ex] \hline
\end{tabular}
\caption{ Non-vanishing $a^4_{mn}$} 
\end{table}
\end{center}

\begin{center}
\begin{table}[h]

\begin{tabular}{|l |lr|}     
\hline

&&\\
~~$\dsp b^{1}_{21}$~~
& $\dsp\frac{1}{48}\,\beta s_{\ell}\,s_{K} \sqrt{(x^2-1)}\Big(-24\,\Re(AB^{\star})-\beta^2((x^2-1)\,\Re(BF^{\star}) +2x\,\Re(BE^{\star})$
&\\& $ +3(x^2-1)\,\Re(CE^{\star}))\Big)$
&\\ 

&&\\  
~~$\dsp b^{1}_{22}$~~
& $\dsp\frac{1}{96}\,\beta^2 s_{\ell}\,s_{K}\Big(24|B|^2 x+|E|^2 x(x^2-1)\,\beta^2-12(x^2-1)\,\Re(AE^{\star})$
 &\\& $+24(x^2-1)\,\Re(BC^{\star})+(x^2-1)^2\,\beta^2\,\Re(FE^{\star})\Big) $
&
\\  [2ex]
&&\\ 
~~$\dsp b^{1}_{23}$~~
& $ \dsp\frac{1}{16}\,\beta^3 s_{\ell}\,s_{K} \sqrt{(x^2-1)}\Big((x^2-1)\,\Re(BF^{\star})+2x\Re(BE^{\star})+(x^2-1)\,\Re(CE^{\star})\Big)$
&
 \\ [2ex] 
&&\\
~~$\dsp b^{1}_{24}$~~
&$\dsp\frac{1}{64}\,\beta^4 s_{\ell}\,s_{K}(x^2-1)\Big(|E|^2x+(x^2-1)\,\Re(FE^{\star}) \Big)$
&
\\ [2ex] \hline
\end{tabular}
\caption{ Non-vanishing $b^1_{mn}$} 
\label{table5}
\end{table}
\end{center}

\begin{center}
\begin{table}[h]

\begin{tabular}{|l |lr|}     
\hline
&&\\
~~$\dsp b^{3}_{21}$~~
& $ \dsp\frac{1}{48}\,\beta s_{\ell}\,s_{K}(x^2-1)\Big(-24\,\Im(AD^{\star})+\beta^2(3\,x\,\Im(BG^{\star})$
&\\& $+(x^2-1)(3\,\Im(CG^{\star}) +\,\Im(DF^{\star})) +x\,\Im(DE^{\star}))\Big)$
&
\\ [2ex]
&&\\ 
~~$\dsp b^{3}_{22}$~~
& $\dsp\frac{1}{96}\,\beta^2 s_{\ell}\,s_{K}\sqrt{(x^2-1)}\Big(-12(x^2-1)\,\Im(AG^{\star})+24x\,\Im(BD^{\star})$
&\\& $+(x^2-1)(24\,\Im(CD^{\star})+\,\beta^2((x^2-1)\,\Im(FG^{\star}) +x\,\Im(EG^{\star})))\Big)$
&\\ [2ex]

&&\\
~~$\dsp b^{3}_{23}$~~
& $\dsp\frac{1}{16}\,\beta^3 s_{\ell}\,s_{K}(x^2-1)\Big(x\,\Im(BG^{\star})+(x^2-1)(\Im(CG^{\star})-\Im(DF^{\star}))-x\Im(DE^{\star})\Big) $
&
\\ [2ex]
&&\\  
~~$\dsp b^{3}_{24}$~~
& $\dsp\frac{1}{64}\,\beta^4 s_{\ell}\,s_{K}(x^2-1)^{3/2}\Big((x^2-1)\,\Im(FG^{\star})+x \,\Im(EG^{\star})$

&\\[2ex]  
\hline
\end{tabular}
\caption{ Non-vanishing $b^3_{mn}$} 
\label{table6}
\end{table}
\end{center}
\end{document}